# Direct momentum-resolved observation of one-dimensional confinement of externally doped electrons within a single subnanometre-scale wire

*Inkyung Song[1]\*\*, Dong-Hwa Oh[1]\*\*, Ha-Chul Shin[1], Sung-Joon Ahn[1], Youngkwon Moon[1], Sun-Hee Woo[2], Hyoung Joon Choi[3], Chong-Yun Park[1,4]\*, and Joung Real Ahn[1,5]\**

[1]Department of Physics, Sungkyunkwan University, Suwon 440-746, Republic of Korea,

[2]College of Pharmacy, Chungnam National University, Daejeon 305-764, Republic of Korea,

[3]Department of Physics, Yonsei University, Seoul 120-746, Republic of Korea,

[4]Department of Energy Science, Sungkyunkwan University, Suwon 440-746, Republic of Korea,

[5]SAINT, Sungkyunkwan University, Suwon 440-746, Republic of Korea

# ABSTRACT

Cutting-edge research in the band engineering of nanowires at the ultimate fine scale is related to the minimum scale of a nanowire-based device. The fundamental issue at the subnanometre-scale is whether angle-resolved photoemission spectroscopy (ARPES) can be used to directly measure the momentum-resolved electronic structure of a single wire because of the difficulty associated with assembling single wire into an ordered array for such measurements. Here, we demonstrated that the one-dimensional (1D) confinement of electrons, which are transferred from external dopants, within a single subnanometre-scale wire (subnanowire) could be directly measured using ARPES. Convincing evidence of 1D electron confinement was obtained using two different gold subnanowires with characteristic single metallic bands that were alternately and spontaneously ordered on a stepped silicon template, Si(553). Noble metal atoms were adsorbed at room temperature onto the gold subnanowires while maintaining the overall structure of the wires. Only one type of gold subnanowires could be controlled using external noble metal dopants without transforming the metallic band of the other type of gold subnanowires. This result was confirmed by scanning tunnelling microscopy experiments and first-principles calculations. The selective control clearly showed that externally doped electrons could be confined within a single gold subnanowire. This experimental evidence was used to further investigate the effects of the disorder induced by external dopants on a single subnanowire using ARPES.

Band engineering is the primary means of imposing desired electronic properties on single nanowires in fundamental research and applications[1-6]. At the subnanometre scale, a single wire resembles a one-dimensional (1D) system, whereas the use of external dopants in band engineering requires a different regime from wires with nanometre-scale widths. For example, the external dopants themselves can produce strong disorder in a single subnanowire, as described by Anderson localisation[7-9]. Furthermore, the motion of electrons along only one pathway may further enhance this disorder, as is known to occur in Luttinger liquids[10,11]. The strong disorder induced by external dopants tends to break single pristine subnanowires; however, the use of external dopants is essential for band engineering. Exotic phenomena associated with 1D band engineering at the subnanometre scale have been studied experimentally by measuring momentum-integrated (k-integrated) electronic structures using field effect transistors or by *k*-integrated tunnelling spectroscopy[5,6]. However, while the use of ARPES is critical for measuring the *k*-resolved electronic structures of single subnanowires, experimental limitations are encountered with using this technique for single subnanowires. For subnanometre layers such as graphene, ARPES can be used to both identify a single subnanometre layer and measure its *k*-resolved electronic structure[12,13]. However, ARPES cannot be used to examine a single subnanowire because the photon beam cannot be focused down to the subnanometre scale; thus, identical subnanowires must be regularly aligned along a specific direction on a surface (Figs. 1a and b). Single subnanowires that do not exhibit atomic coupling between the wires, such as carbon nanotubes, cannot be aligned regularly on a surface: thus, subnanowires that exhibit atomic coupling, such as self-assembled subnanowires on a surface, are more suitable for this study. Subnanowires with atomic coupling can be aligned on a surface, but the electronic coupling between these wires presents its own challenge. Thus, it is not obvious whether electronic structural changes occur only within a single subnanowire or if the electronic coupling also affects the neighbouring subnanowires (see Supplementary Fig. S1).

To resolve these issues, subnanowires must be regularly aligned by atomic coupling. In addition, the electronic coupling must be significantly weaker than the atomic coupling. Self-assembled gold subnanowires on a Si(553) surface, which is a vicinal Si(111) surface, were used in this study[14-21]. Two different types of metallic gold subnanowires were located on a single terrace and were regularly and alternately aligned across the step edge direction. Each gold subnanowire produced a characteristic single metallic energy band[18]. The ARPES measurements for these metallic energy bands were used to identify the two gold subnanowires (see Figs. 1b-d). When only one type of metallic subnanowire was aligned (see Supplementary Fig. S1), ARPES could not be used to clearly determine whether the externally doped electrons were transferred across wires (see Supplementary Fig. S1b) or confined within a single subnanowire (see Supplementary Fig. S1c). Direct observations by ARPES measurements (see Figs. 1e-g) showed that adsorbing extra noble metal atoms onto self-assembled gold subnanowires at room temperature (RT) shifted only one of the two metallic energy bands to a higher binding energy, leaving the other metallic energy band unchanged. This result clearly demonstrates that only one type of metallic gold subnanowire could be selectively controlled by the use of external dopants. When a strong electronic coupling between the two types of metallic gold subnanowires exists (see Fig. 1c), the two metallic energy bands are expected to shift simultaneously. Therefore, the ARPES measurements showed that the two different types of gold subnanowires were electronically decoupled at RT, despite being atomically coupled. This suggests that the ARPES measurement of a single subnanowire can be realised by the spontaneous assembly of multiple types of subnanowires with different atomic and electronic coupling strengths.

The highly anisotropic structures of vicinal Si(111) surfaces have been used as templates in the self-assembly of subnanowires. Among these vicinal Si(111) surfaces, Si(553) and Si(557) surfaces have primarily been used to investigate exotic 1D phenomena in self-assembled subnanowires[14-27]. In this study, self-assembled gold subnanowires on a Si(553) surface were used as a pristine system, where the Si(553) surface exhibited a miscut crystallographic angle of 12.5° towards [11-2] from the [111] orientation[14-21]. Mariusz Krawiec *et al.*[18] developed an updated atomic structure model in which the gold-covered Si(553) surface is composed of two types of gold subnanowires. They coexist within a single terrace with a width of 1.48 nm and are aligned alternately across the step edge direction (see Supplementary Fig. S2). Each gold subnanowire has a characteristic single metallic energy band[18]. The metallic energy band of one of the two gold subnanowires ($S_1$ in Fig. 1e) is nearly half filled, whereas the metallic energy band of the other gold subnanowire ($S_2$ in Fig. 1e) is nearly quarter-filled[14-16]. The $S_1$ energy band is split by the spin-orbit coupling of the hybridised gold-silicon orbitals ($S_1{'}$, $S_1$)[22,23]. This electronic structure makes these two types of gold subnanowires highly appropriate for this study. Note that the prediction of coexistence between two different gold subnanowires in a single terrace will not change with subsequent modifications to the updated atomic structure model because the gold coverage has been experimentally confirmed.

Extra gold atoms were introduced as external dopants at RT to control the electronic band structures of the subnanowires. Increasing the amount of extra gold atoms shifted the metallic $S_2$ energy band to a higher binding energy, whereas the metallic $S_1$ energy band was robust to the change (see Figs. 1e-g). The changes in the energy bands clearly showed that the extra gold atoms acted as external electron dopants and that the electrons transferred from the gold dopants were delocalised in only one type of gold subnanowires. More precisely, the changes in the energy bands showed that the externally doped electrons were confined within a single gold subnanowire (see Fig. 1d). This selective control of multiple metallic (or dispersive) energy bands is very unique. For example, external sodium dopants have been used to simultaneously shift the three metallic energy bands of indium subnanowires on a Si(111) surface, and external silicon dopants have been used to simultaneously shift the two metallic energy bands of gold subnanowires on a Si(111) surface[28,29]. To determine whether the extra gold atoms induced 1D delocalisation within a single gold subnanowire, silver atoms were introduced in the place of the extra gold atoms on the pristine gold subnanowires at RT (see Supplementary Fig. S3). The change in the electronic structure induced by the silver atoms was very similar to that induced by the extra gold atoms. This result showed that the 1D delocalisation within a single gold subnanowire was not specific to the extra gold atoms.

The ARPES experiments directly showed that the extra gold dopants induced a shift in the metallic $S_2$ energy band within a single gold subnanowire. These concrete experimental results for the $k$-resolved electronic structure motivated further studies on the effects of disorder on 1D electron delocalisation within a single subnanowire: these effects are important in 1D band engineering applications at the ultimate nanoscale level and in fundamental research, such as that on disordered Luttinger liquids. To investigate the effects of disorder on the electronic structure induced by extra gold atoms in detail, the momentum distribution curves (MDCs) and energy distribution curves (EDCs) of the gold subnanowires were measured using increasing amounts of extra gold atoms (see Figs. 2a-c). The three Fermi momentums ($k_F$'s) of the MDC arose from the spin-orbit-split $S_1'$, $S_1$ and $S_2$ energy bands located at -1.27, -1.238, and -1.046 $\text{Å}^{-1}$, respectively. Increasing the amount of extra gold atoms gradually moved the $k_F$ of the $S_2$ energy band towards a higher momentum until there were 0.048 monolayers (ML) of extra gold coverage. Above 0.048 ML of extra gold coverage, the $k_F$ of the $S_2$ energy band saturated at -1.104 $\text{Å}^{-1}$, whereas the $k_F$ of the $S_1$ energy band was maintained regardless of the amount of the extra gold atoms. Figure 2d shows the changes in the electron filling of the $S_1$ and $S_2$ energy bands, as determined from the $k_F$ values, for which the electron filling of the $S_2$ energy band increased from 0.276 to 0.347, whereas the electron filling of the $S_1$ energy band remained at 0.51. The EDC at each $k_F$ of the $S_2$ energy band was generated to determine its metallic properties as a function of the amount of extra gold atoms (see Fig. 2c). The $S_2$ energy band was metallic before the $k_F$ saturated for the $S_2$ energy band (see Fig. 2e). Above the critical amount of extra gold coverage at which the $k_F$ saturated, a finite density of states at $E_F$ remained; however, a very small energy gap, known as the pseudo energy gap, began to open (see Fig. 2e). This result showed that there was a limit to the external electron doping by the extra gold atoms. Thus, the extra gold atoms did not act as external electron donors above this critical doping level, instead disrupting the intrinsic electronic structure of the pristine gold subnanowires. The extra gold atoms inevitably induced local disorder, even though the overall atomic structure was maintained. The degree of disorder introduced by the extra gold atoms was determined from the full width at half maximum (FWHM) of the MDC peak, which was inversely proportional to the coherence length along the subnanowire. The relationship between the FWHM and the coherence length of the $S_2$ energy band is shown in Fig. 2f. Here, we note that the FWHM of the MDC peak can be further broaden by other effects so that the calculated coherent length in Fig. 2f can be smaller than an intrinsic coherence length. First, the spectrometer has a limited momentum resolution. Second, the $S_2$ energy band also could be spin-split although the splitting was not resolved in ARPES experiments because the FWHM of the $S_2$ energy band is twice that of the spin-split $S_1$ and $S_1$' energy bands, as reported in other ARPES experiment[14]. Third, intrinsic defects on the pristine gold-covered Si(553) surface have an influence on subnanowires up to 20 nm, depending on types of defects and temperature[30]. The FWHM of the $S_2$ energy band increased and the coherence length decreased as the amount of extra gold atoms increased. The change in the FWHMs showed that disorder accumulated continuously along the subnanowire as the amount of extra gold atoms increased. When silver atoms were used instead of extra gold atoms, very similar changes in the MDCs and EDCs were observed (see Supplementary Fig. S3). Disorder can significantly affect transitions in a 1D electron system, as observed in Anderson localisation and Luttinger liquids[7-11]. Disorder tends to localise electrons in a 1D system: above a critical disorder, this localisation can induce a 1D metal-insulator transition. The ARPES measurements showed that external dopants could be used to dope a single subnanowire up to a critical disorder and that the amount of induced disorder could be used to control the 1D metal-insulator transition.

Scanning tunnelling microscopy (STM) images were obtained (see Fig. 3) with increasing amounts of extra gold atoms to better understand the process of selective band engineering. Figure 3a shows an STM image of pristine gold subnanowires on the Si(553) surface. The updated atomic structure model was used to demarcate the STM image with white lines originating from the Si step edge structure: these white lines defined the terraces within which the two gold subnanowires were located[18]. After adsorption of 0.004 ML of extra gold atoms, double protrusions were observed on the Si step edge structure (see Figs. 3b-d). The size of a single protrusion was 0.73 nm (Fig. 3f), which was comparable to the size of a single gold atom[31,32]. The number of identical protrusions increased with the amount of extra gold atoms. At a lower bias voltage ($V_s$) of -0.5 V, modulations with a period of $2a_0$ were observed on the terrace, where $a_0$ denotes the unit atomic distance along the wire; these modulations were not clearly observable at a higher $V_s$ of -1.0 V (see Figs. 3e and f). The $2a_0$ modulations were rarely found on the pristine gold subnanowires at RT but were observed more frequently as the amount of extra gold atoms increased[30]. Thus, the $2a_0$ modulations suggest that extra gold atoms may also have been adsorbed on the terrace[33].

The reversible control of the 1D electron doping was also investigated (see Fig. 4). Annealing the sample at 650 °C following the adsorption of extra gold atoms at RT resulted in the complete recovery of the electronic band structure of the pristine gold subnanowires (see Figs. 4d and f). This result showed that a simple thermal treatment could be used to selectively remove the extra gold atoms from the gold subnanowires, thereby recovering the atomic structure of the pristine gold subnanowires. When the extra gold atoms were adsorbed again, the electronic band structure changed to an electron-doped structure. The reversible processes between the undoped and doped electronic structures were successfully repeated over several cycles. The reversible electronic structure control of the undoped and doped gold subnanowires was confirmed by the STM images (see Figs. 4a, c, and e). After thermal annealing at 650 °C, the protrusions in the STM image disappeared, and the atomic structure of the pristine gold subnanowires was recovered.

First-principles calculations were also used to investigate the underlying atomic structure of gold subnanowires covered by extra gold atoms. The atomic structure model for pristine gold subnanowires developed by Mariusz Krawiec *et al.*[18] is shown in Fig. 5a and Supplementary Fig. S2. First, the surface free energies were calculated for various possible adsorption sites, corresponding to an inter-atom distance of $2a_0$ between the extra gold atoms (see Supplementary Figs. S4a-f), as was observed in STM images. The lowest surface free energy was found for extra gold atoms adsorbed near the pristine gold subnanowire (see Supplementary Figs. S4d and k). The second lowest surface free energy was found for extra gold atoms located at the end of the silicon step edge structure (see Supplementary Figs. S4a and k), at the corner of the silicon step edge structure (see Supplementary Figs. S4b and k), or on the silicon step edge structure (see Supplementary Figs. S4c and k). The surface free energies increased (see Supplementary Fig. S4k) for inter-atom distances of $a_0$ between the extra gold atoms in the four models (see Supplementary Figs. S4g-j). This result showed that the extra gold atoms were repulsive independent of the adsorption site, which was supported by the observation of extra gold atoms with an inter-atomic distance of $2a_0$. Simulated STM images of the various adsorption sites were also obtained (see Fig. 5 and Supplementary Fig. S5). Most of the features of the experimental STM images were reproduced in the simulated STM images when the extra gold atoms were located at the end of the silicon step edge structure (see Fig. 5b) or near the pristine gold subnanowires (see Fig. 5c). The simulated STM images showed that the bright protrusions separated by $2a_0$ (both at the silicon step edge and on the terrace) could be ascribed to the extra gold atoms. The surface free-energy calculations and the simulated STM images showed that extra gold atoms could be found both at the ends of the silicon step edge structures and near the pristine gold subnanowires[33]. The surface free-energy results also show that more extra gold atoms could be adsorbed near the pristine gold subnanowires relative to those at the end of the silicon step edge structures.

Electronic band structures were calculated for the gold subnanowires doped with extra gold atoms (see Fig. 5). The unfolded electronic band structures were calculated to compare with the experimental energy bands (see Figs. 5e-h). The electronic band structure for the pristine gold subnanowires (shown in red in Figs. 5e-h) was juxtaposed with those for the gold subnanowires doped with extra gold atoms (shown in grey in Figs. 5f-h). Adsorbing extra gold atoms at the end of the silicon step edge structure did not shift the $S_1$ and $S_2$ energy bands (shown in grey in Fig. 5f) relative to the $S_1$ and $S_2$ energy bands of the pristine gold subnanowires (shown in red in Fig. 5f). In contrast, when extra gold atoms were adsorbed near the pristine gold subnanowires, only the $S_2$ energy band shifted towards the higher energy band by approximately 0.25 eV (see Fig. 5g), as was observed in the ARPES experiment. When extra gold atoms were adsorbed both at the end of the silicon step edge structures and near the pristine gold subnanowires, the electronic band structure was similar to that found when extra gold atoms were adsorbed only near the pristine gold subnanowires (see Fig. 5h). The predicted electronic band structures showed that the extra gold atoms at the end of the silicon step edge structure did not affect the $S_1$ and $S_2$ energy bands and that the selective control of the $S_2$ energy band could be ascribed to the extra gold atoms near the pristine gold subnanowires. The $S_1$ and $S_2$ energy bands originate from different gold subnanowires, based on the atomic structure model of Mariusz Krawiec et al.[18], as described above. The $S_1$ energy band originates from the $Au_1$ and $Si_1$ atoms and the $S_2$ energy band comes from the $Au_2$ and $Si_2$ atoms (see Fig. 5a). When extra gold atoms were located near the $Au_2$ atom, which is the opposite side of the $Au_1$ atom, the selective electron doping was reproduced. The results suggest that, based on the atomic structure model, the extra gold atoms transfer electrons to the $Au_2$ atoms but the transferred electrons are not further delocalised though the $Au_1$ atoms. The selective electron doping may be because of weak electronic interactions between the $Au_1$ and $Au_2$ atoms or the energetic instability of the $S_1$ energy band for electron doping, compared to the energetic stabilization of the $S_2$ energy band by electron doping, or thermodynamic effects at RT, as described in the following. In the ARPES experiment, 0.048 ML of extra gold atoms shifted the $S_2$ energy band by approximately 0.15 eV (see Fig. 1g). In the calculations, adsorbing 0.1 ML of extra gold atoms near the pristine gold subnanowires shifted the $S_2$ energy band by approximately 0.25 eV, which was not significantly different from the ARPES results. The small difference between the predicted and experimental results may be attributed to RT thermodynamic effects, such as thermal fluctuations of the pristine subnanowires, the thermal reduction in the electronic and atomic coupling between the subnanowires, and the diffusion of extra gold atoms on the terrace, which cannot be completely accounted for in first-principles calculations performed at absolute zero temperature[33,34]. Furthermore, the RT thermodynamic effects may results in much obvious selective control of the $S_1$ energy band, compared to the theoretical calculations. On the other hand, the pristine gold-covered Si(553) surface was reported to undergo a metal-insulator transition at a low temperature, where the origin of the metal-insulator transition was suggested to be Peierls instability[16,35]. At a low temperature, the gold subnanowire on the terrace and the silicon step edge structure show $2a_0$ and $3a_0$ modulations, respectively. In the concept of Peierls instability, the temperature-dependent metal-insulator transition is order-order. In contrast to the temperature-dependent metal-insulator transition, the metal-insulator transition induced by extra gold atoms at RT is order-disorder. The extra gold atoms do not show a long-range order. The randomly-located extra gold atoms result in local $2a_0$ modulations of gold subnanowires on the terrace, but do not induce a $3a_0$ modulation on the silicon step edge structures.

In summary, the *k*-resolved electronic structure associated with the 1D confinement of electrons was directly observed using ARPES experiments. The 1D confinement was observed using external electron doping to selectively control two characteristic metallic energy bands of gold subnanowires and was confirmed by first-principles calculations. The selective control was performed reversibly using repetitive cycles of adsorption of extra gold atoms at RT and thermal treatment. The *k*-resolved electronic structure was used to study the effects of disorder on 1D confinement and revealed a critical disorder that divided a 1D delocalisation regime from a 1D localisation regime along the wire. This interesting 1D electron confinement at widths on subnanometre scales can be extended to describe exotic 1D phenomena such as 1D Anderson localisation, Luttinger liquids, and disordered Luttinger liquids using ARPES experiments.

## METHODS

### Experiment

A Si(553) surface was cleaned by direct resistive heating. Gold subnanowires were grown by depositing gold on the Si surface while maintaining the sample at 650 ℃, followed by annealing at 850 ℃. Angle-resolved photoemission spectroscopy (ARPES) measurements were performed using a high-resolution electron analyser (SIENTA R3000, Gamma Data) equipped with a high-flux monochromator for He II radiation ($hv = 40.2$ eV). A commercial low-temperature STM (LT-STM, Omicron) was used to obtain the STM images. The ARPES spectra and the STM images were obtained at RT.

### Theory

First-principles calculations using density functional theory were carried out using the Vienna *Ab initio* Simulation Package (VASP) code[36]. In these calculations, the potential between an electron and an ion was described using the projected augmented wave method. The exchange-correlation part of the electron-electron interactions was described using the generalised gradient approximation of Perdew *et al*[37-39]. Plane waves were used as the basis functions for the electronic wave functions with a kinetic energy cut-off of 250 eV. The calculations were performed using a slab geometry composed of three Si bilayers stacked along the [111] direction, the bottom layer of which was saturated by hydrogen atoms. The slab was separated by approximately 10 Å of vacuum. The total energy was minimised using quenched dynamics and quasi-Newtonian methods, with a criterion of force of 0.02 eV/Å . The STM images were simulated using the Tersoff and Hamann method[40]. The unfolded energy bands were calculated using H. J. Choi's method[41].

**ASSOCIATED CONTENT**

**Supporting Information**. Schematic of angle-resolved photoemission spectroscopy configuration, atomic structure model of the pristine gold subnanowires suggested by Mariusz Krawiec *et al*., selective band engineering of multiple metallic energy bands with silver atoms, atomic structure models of the gold subnanowires with extra gold atoms, and simulated STM images of the pristine gold subnanowires with extra gold atoms. This material is available free of charge via the Internet at http://pubs.acs.org.

**AUTHOR INFORMATION**

**Corresponding Author**

*E-mail: cypark@skku.edu (C.-Y. Park), jrahn@skku.edu (J. R. Ahn)

**Author Contributions**

**These authors contributed equally to this work.

**Notes**

The authors declare no competing financial interest.

**ACKNOWLEDGMENT**

This study was supported by a National Research Foundation of Korea (NRF) grant funded by the Korea government (MEST) (No. 2012R1A1A2041241) and and the National Research Foundation (NRF) of Korean grant funded by the Korean Ministry of Science, ICT and Planning (No. 2012R1A3A2048816).

**REFERENCES**

1. Weber, B.; Mahapatra, S.; Ryu, H.; Lee, S.; Fuhrer, A.; Reusch, T. C. G.; Thompson, D. L.; Lee, W. C. T.; Klimeck, G.; Hollenberg, L.C. L.; Simmons, M. Y.. Science **2012**, 335, 64-67.

2. Gudiksen, M. S.; Lauhon, L. J.; Wang, J.; Smith, D. C.; Lieber, C. M. Nature **2002**, 415, 617-620.

3. Yang, C.; Zhong, Z.;  Lieber, C. M. Science **2005**, 310, 1304-1307.

4. Ho, J. C.; Yerushalmi, R.; Jacobson, Z. A.; Fan, Z.; Alley, R. L.; Javey, A. Nat. Mater. **2008**, **7**, 62-67.

5. Zhou, C.; Kong, J.; Yenilmez, E.; Dai, H. Science **2000**, 290, 1552-1555.

6. Samarajeewa, D. R.; Dieckmann, G. R.; Nielsen, S. O.; Musselman, I. H. Carbon **2013**. 57, 88.

7. Anderson, P. W. Phys. Rev. **1958**, 109, 1492-1505.

8. Gómez-Navarro, C.; De Pablo, P. J.; Gómez-Herrero, J.; Biel, B.; Garcia-Vidal, F. J.; Rubio, A.; Flores, F. Nat. Mater. **2005**, 4, 534-539.

9. Dietl, P.; Metalidis, G.; Golubev, D.; San-Jose, P.; Prada, E.; Schomerus, H.; Schön, G. Phys. Rev. B **2009**, 79, 195413.

10. Bockrath, M.; Cobden, D. H.; Lu, J.; Rinzler, A. G.; Smalley, R. E.; Balents, L.; McEuen, P. L. Nature **1999**, 397, 598.

11. Deshpande, V. V.; Bockrath, M.; Glazman, L. I.; Yacoby, A. Nature **2010**, 464, 209.

12. Geim, A. K. Science **2009**, 324, 1530.

13. Emtsev, K. V.; Bostwick, A.; Horn, K.; Jobst, J.; Kellogg, G. L.; Ley, L.; McChesney, J. L.; Ohta, T.; Reshanov, S. A.; Röhrl, J.; Rotenberg, E.; Schmid, A. K.; Waldmann, D.; Weber, H. B.; Seyller, T. Nat. Mater. **2009**, 8, 203-207.

14. Crain, J. N.; Kirakosian, A.; Altmann, K. N.; Bromberger, C.; Erwin, S. C.; McChesney, J. L.; Lin, J.-L.; Himpsel, F. J. Phys. Rev. Lett. **2003**, 90, 176805.

15. Crain, J. N.; McChesney, J. L.; Zheng, F.; Gallagher, M. C.; Snijders, P. C.; Bissen, M.; Gundelach, C.; Erwin, S. C., Himpsel, F. J. Phys. Rev. B **2004**. 69, 125401.

16. Ahn, J. R.; Kang, P. G.; Ryang, K. D.; Yeom, H. Y. Coexistence of two different Peierls distortions within an atomic scale wire: Si(553)-Au. Phys. Rev. Lett. **2005**, 95, 196402.

17. Snijders, P. C.; Rogge, S.; Weitering, H. H. Phys. Rev. Lett. **2006**, 96, 076801.

18. Krawiec, M. Phys. Rev. B **2010**, 81, 115436.

19. Voegeli, W.; Takayama, T.; Shirasawa, T.; Abe, M.; Kubo, K.; Takahashi, T.; Akimoto, K.; Sugiyama, H. Phys. Rev. B **2010**, 82, 075426.

20. Crain, J. N.; Pierce, D. T. Science **2005**, 307, 703-706.

21. Erwin, S. C.; Himpsel, F. J. Nat. Commun. **2010**, 1, 58.

22. Sánchez-Potal, D.; Riikonene, S.; Martin, R. M. Phys. Rev. Lett. **2004**, 93, 146803.

23. Barke, I.; Zheng, F.; Rügheimer, T. K.; Himpsel, F. J. Phys. Rev. Lett. **2006**, 97, 226405.

24. Ahn, J. R.; Yeom, H. W.; Yoon, H. S.; Lyo, I.-W. Phys. Rev. Lett. **2003**, 91, 196403.

25. Yeom, H. W.; Ahn, J. R.; Yoon, H. S.; Lyo, I.-W.; Jeong, H.; Jeong, S. Phys. Rev. B **2005**, 72, 035323.

26. Song, I.; Oh, D.-H.; Nam, J. H.; Kim, M. K.; Jeon, C.; Park, C.-Y.; Woo, S. H.; Ahn, J. R. New J. Phys. **2009**, 11, 063034.

27. Kim, M. K.; Oh, D.-H.; Baik, J.; Jeon, C.; Song, I.; Nam, J. H.; Woo, S. H.; Park, C.-Y.; Ahn, J. R. Phys. Rev. B **2010**, 81, 085312.

28. Choi, W. H.; Kang, P. G.; Ryang, K. D.; Yeom, H. Y. Phys. Rev. Lett. **2008**, 100, 126801.

29. Morikawa, H.; Hwang, C. C.; Yeom, H. Y. Phys. Rev. B **2010**, 81, 075401.

30. Ryang, Kyung-Deuk; Kang, Pil Gyu; Jeong, Sukmin; Yeom, Han Woong Phys. Rev. B **2007**, 76, 205323.

31. Kulawik, M.; Nilius, N.; Freund, H.-J. Phys. Rev. Lett. **2006**, 96, 036103.

32. Schäfer, J.; Blumenstein, C.; Meyer, S.; Wisniewski, M.; Claessen, R. Phys. Rev. Lett. **2008**, 101, 236802.

33. Krawiec, M.; Jałochowski, M. Phys. Rev. **2013**, 87, 075445.

34. Grüner, G. Density waves in solids (Addison-Wesley Publishing Co.). **1994**.

35. Okino, Hiroyuki; Matsuda, Iwao; Hobara, Rei; Hasegawa, Shuji; Kim, Younghoon; Lee, Geunseop Phys. Rev. B **2007**, 76, 035424.

36. Kresse, G.; Furthmüller, J. Comput. Mat. Sci. **1996**, 6, 15.

37. Blochl, P. E. Phys. Rev. B **1994**, 50, 17953.

38. Kresse, G.; Joubert, D. Phys. Rev. B **1999**, 59, 1758.

39. Perdew, J. P.; Chevary, J. A.; Vosko, S. H.; Jackson, K. A.; Pederson, M. R.; Singh, D. J.; Fiolhais, C. Phys. Rev. B **1992**, 46, 6671.

40. Tersoff, J.; Hamann, D. R. Phys. Rev. B **1985**, 31, 805.

41. Lee, H.; Kim, S.; Ihm, J.; Son, Y.-W.; Choi, H. J. Carbon **2011**, 49, 2300.

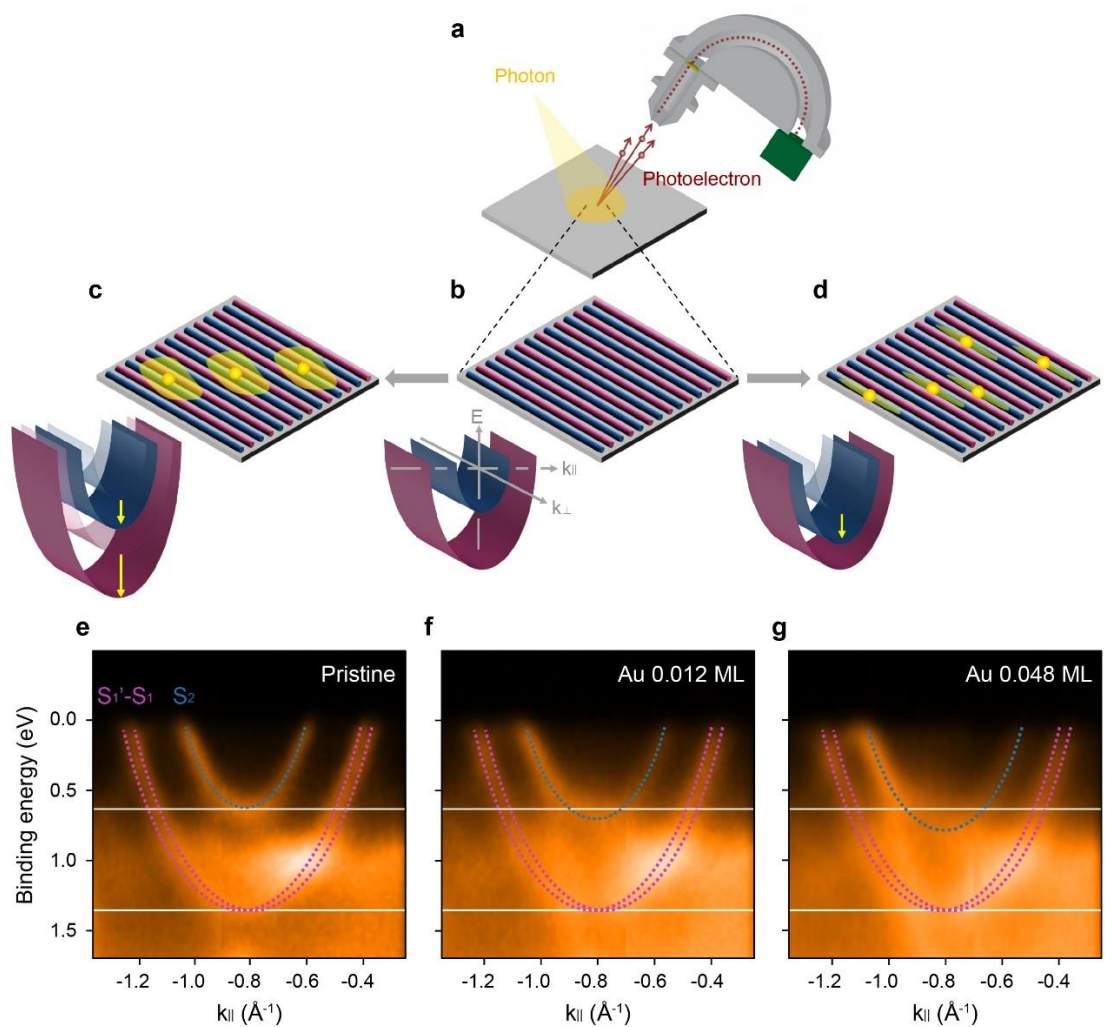

**Figure 1 | Change induced by external dopants in electronic band structures with multiple metallic energy bands:** (**a**) schematic of angle-resolved photoemission spectroscopy configuration; (**b**) schematic showing arrangement of two alternating different subnanowires, represented by red and blue rods, which produce red and blue metallic energy bands, respectively; (**c**) changes induced in metallic energy bands by external dopants (shown as yellow solid spheres) by delocalising externally doped electrons across the subnanowires; (**d**) changes induced in metallic energy bands by external dopants (shown as yellow solid spheres) by localising externally doped electrons within single subnanowires; changes induced in electronic band structures of a gold-covered Si(553) surface (**e**) without extra gold coverage and with

extra gold coverage of **(f)** 0.012 ML and **(g)** 0.048 ML, where the electronic band structures were measured along the wire direction ($k_\parallel$) at RT.

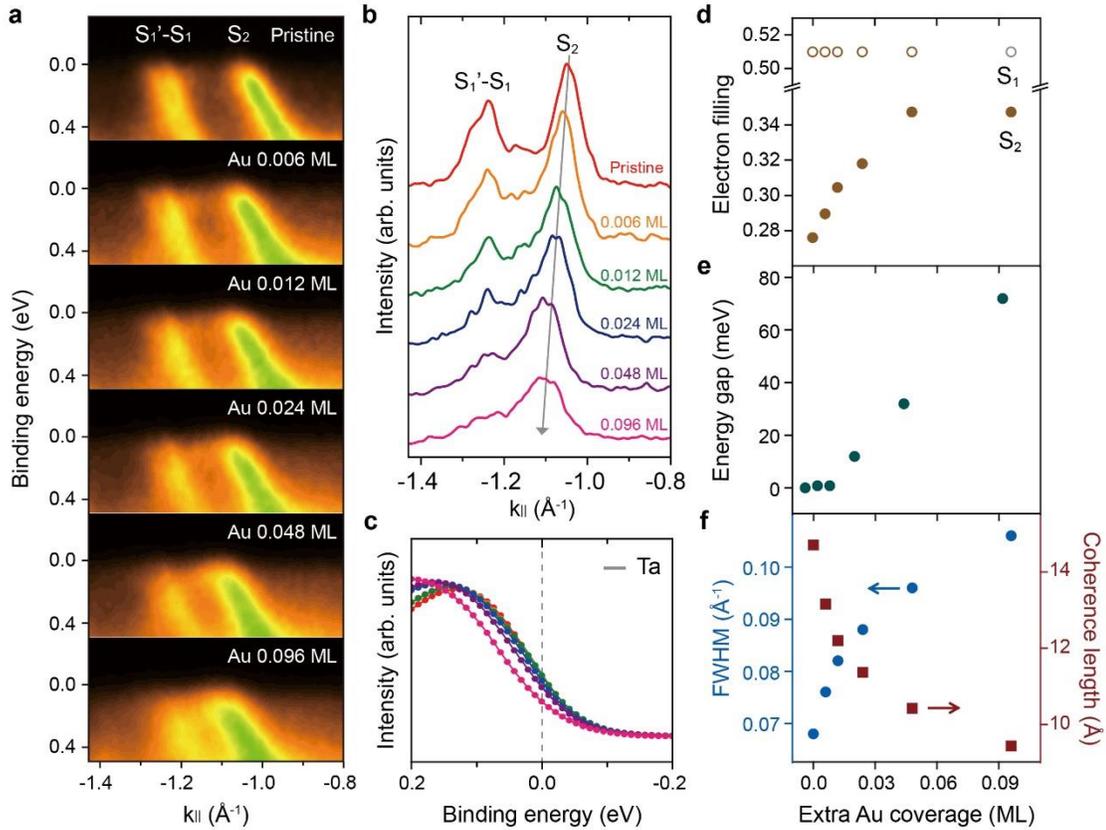

**Figure 2 | Selective band engineering of multiple metallic energy bands with extra gold atoms: (a)** change in the electronic band structure of the pristine surface by increasing the amount of extra gold atoms; **(b)** MDCs of ARPES intensity maps at $E_F$; **(c)** EDCs of ARPES intensity maps at the $k_F$ of the $S_2$ energy band, where the same colours are used in both **(b)** and **(c)** to denote different coverages, and the EDC of a tantalum (shown as a solid grey line) was used to estimate the energy gap depending on the extra gold coverage; **(d)** change in the electron filling of the $S_1$ and the $S_2$ energy bands; **(e)** change in the pseudo energy gap of the $S_2$ energy band; and **(f)** change in the FWHMs of the MDC peaks (blue solid circles) and the coherence length (brown solid rectangles) of the $S_2$ energy band.

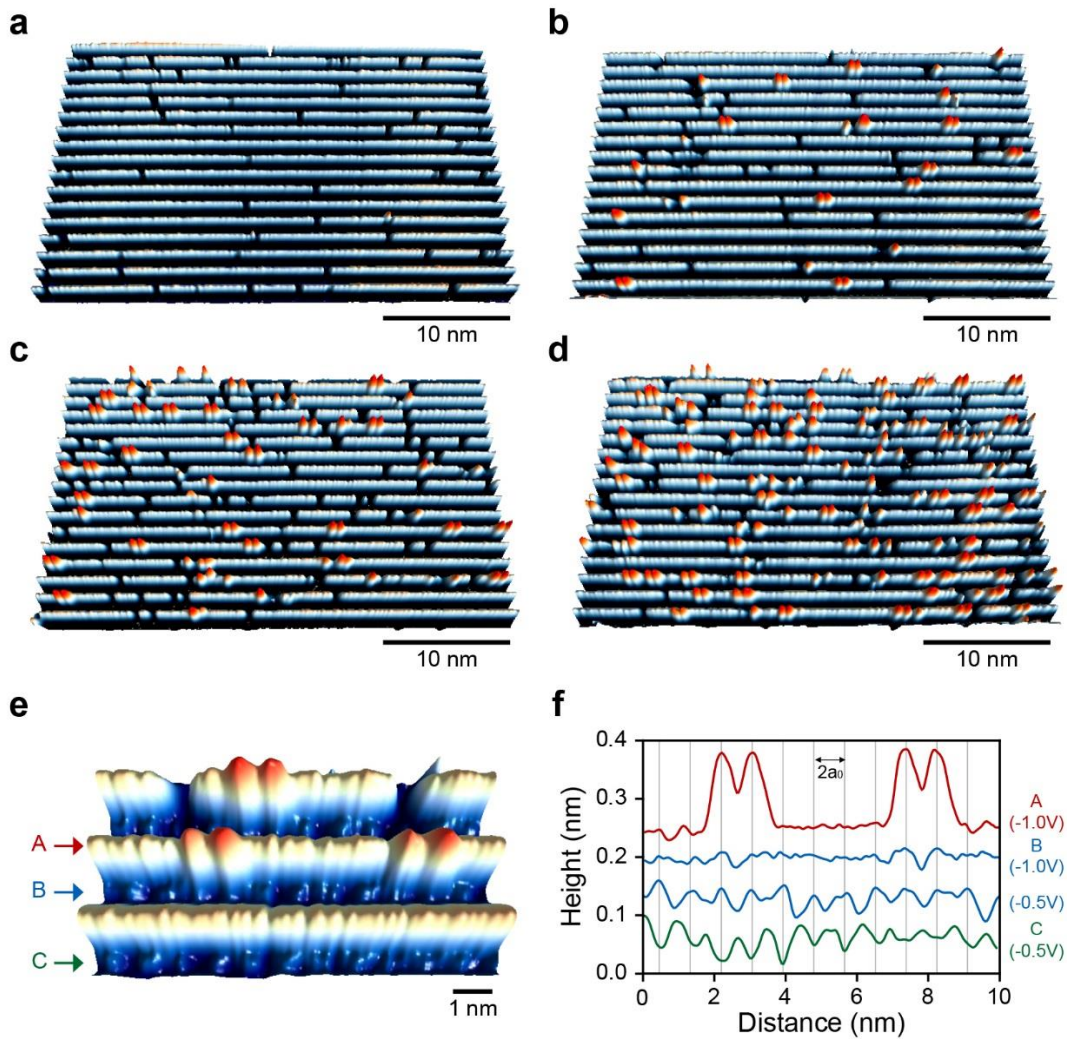

**Figure 3 | Changes in STM images with increasing amounts of extra gold atoms:** STM images of pristine gold subnanowires at RT with extra gold coverages of **(a)** 0 ML, **(b)** 0.004 ML, **(c)** 0.009 ML, and **(d)** 0.02 ML; **(e)** enlarged STM image of pristine gold subnanowires with extra gold atoms, where STM images in **(a)-(d)** were obtained at $V_s$ = -1.0 V, and the STM image in **(e)** was obtained at $V_s$ = -0.5 V; **(f)** line profiles recorded on the silicon step edge (A, red) and on the gold subnanowire (B, blue and C, green) shown in **(e)**.

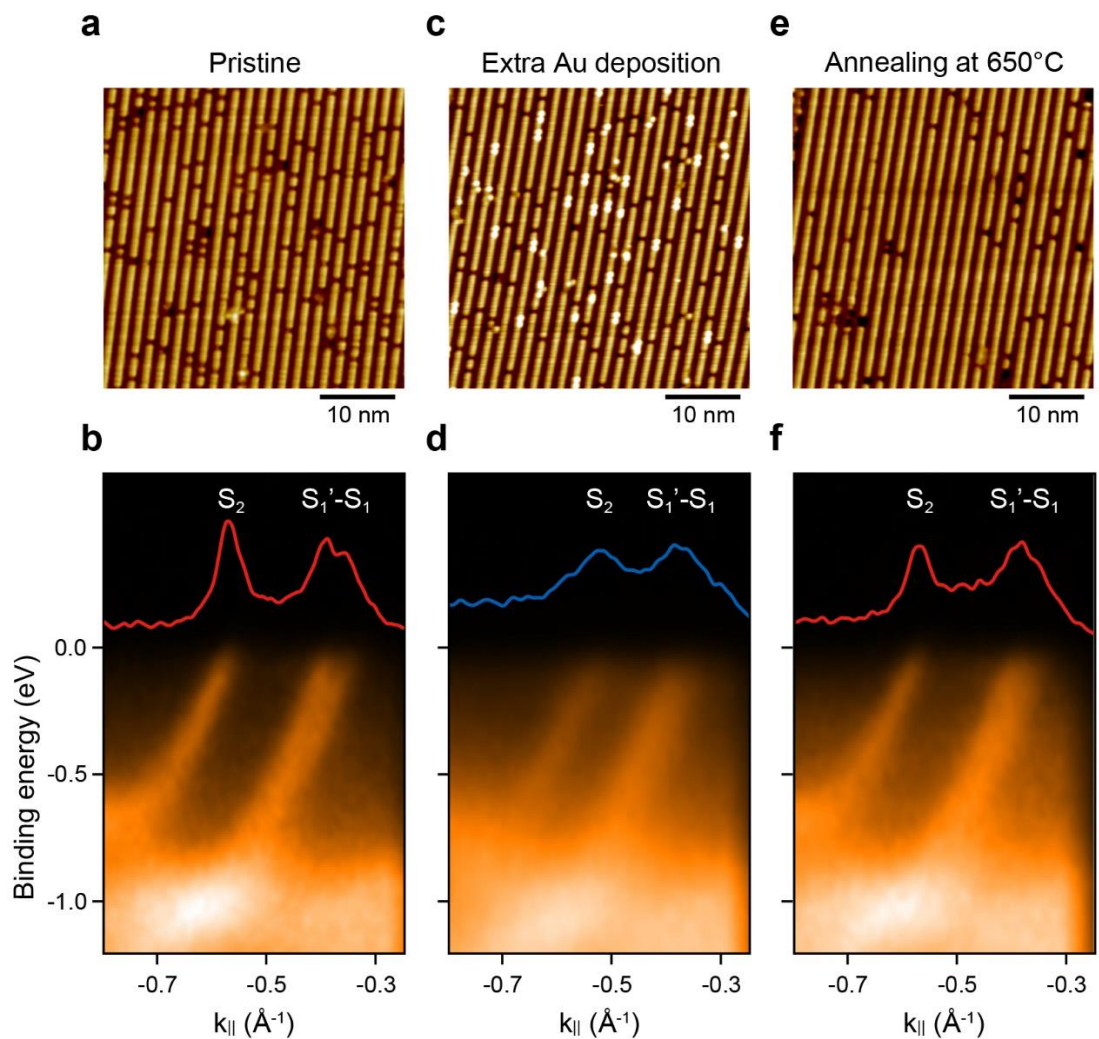

**Figure 4 | Reversible control between undoped and doped gold subnanowires:** STM images and electronic band structures for **(a), (b)** pristine undoped gold subnanowires, **(c), (d)** electron-doped gold subnanowires after the extra gold deposition, and **(e), (f)** undoped gold subnanowires recovered by thermal annealing at 650 ℃, respectively; the upper portions of **(b)**, **(d)**, and **(f)** show the MDCs of the ARPES intensity maps at $E_F$.

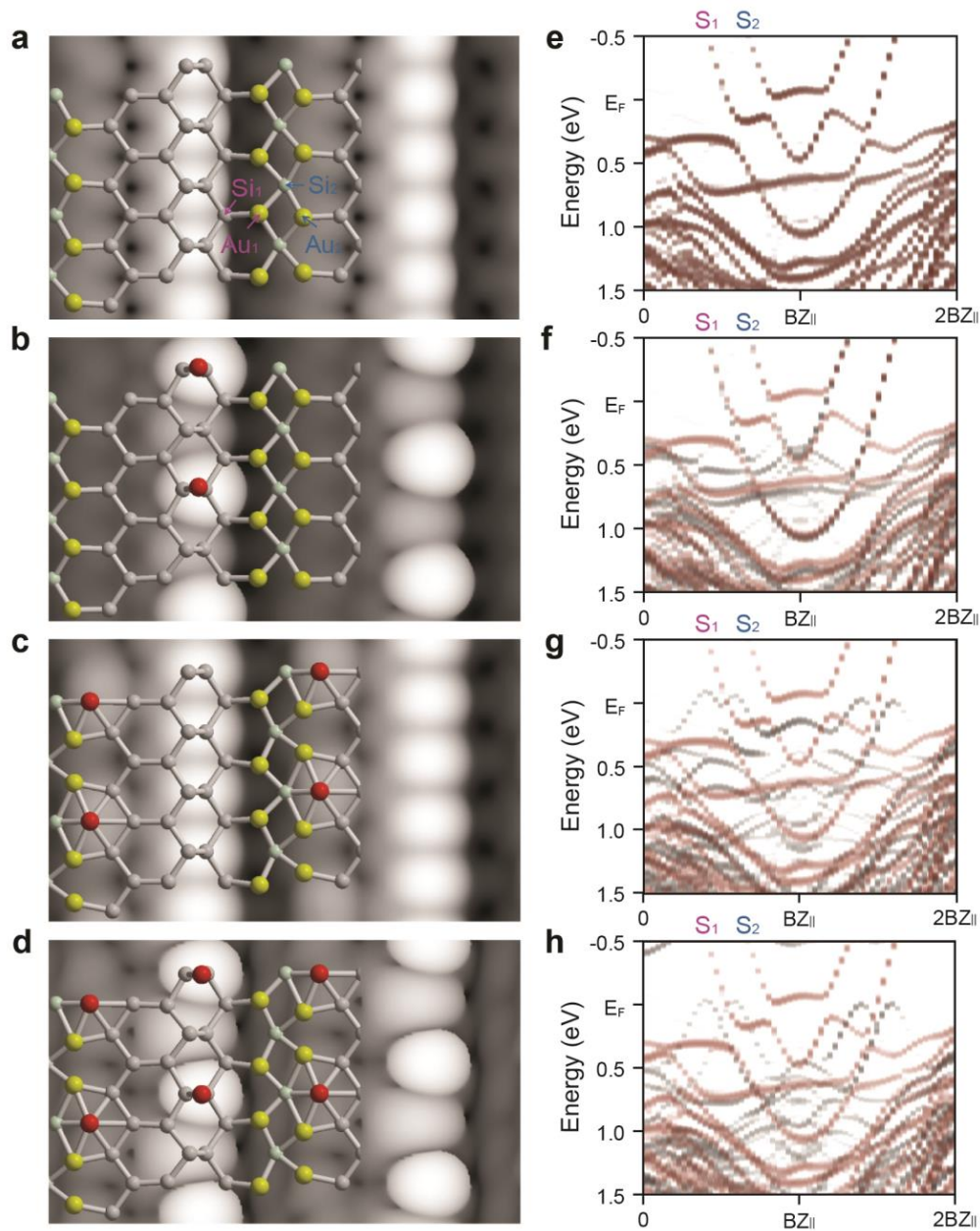

**Figure 5 | Simulated STM images and electronic band structures of gold subnanowires:** simulated STM images using Mariusz Krawiec *et al.*'s atomic structure model **(a)** without extra gold atoms and **(b)-(d)** with extra gold atoms; yellow and grey spheres denote gold and silicon atoms, respectively, and red spheres denote extra gold atoms; the simulated STM images were obtained for an energy window from $E_F$ to $E_F$ -1.0 eV; unfolded electronic band structures calculated using Mariusz Krawiec *et al.*'s atomic structure model **(e)** without extra gold atoms and **(f)-(h)** with extra gold atoms, where the unfolded electronic band structure of the pristine gold subnanowires (shown in red) is juxtaposed with that of the gold subnanowires doped with extra gold atoms (shown in grey).

**TABLE OF CONTENTS**

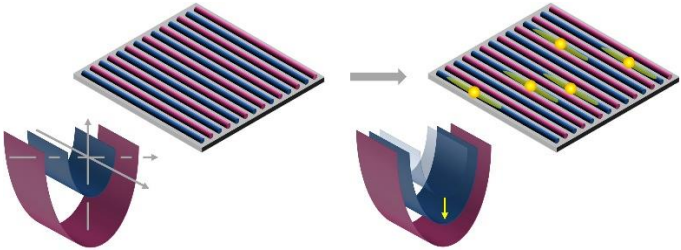